# Mycroft: Tracing Dependencies in Collective Communication Towards Reliable LLM Training

Yangtao Deng[1][*][#], Lei Zhang[2][*][†], Qinlong Wang[3], Xiaoyun Zhi[3], Xinlei Zhang[2],
Zhuo Jiang[2][†], Haohan Xu[2], Lei Wang[2], Zuquan Song[3], Gaohong Liu[3],
Yang Bai[2], Shuguang Wang[3], Wencong Xiao[3], Jianxi Ye[2], Minlan Yu[4], Hong Xu[1]

[1]The Chinese University of Hong Kong   [2]ByteDance   [3]ByteDance Seed   [4]Harvard University

## Abstract

Reliability is essential for ensuring efficiency in LLM training. However, many real-world reliability issues remain difficult to resolve, resulting in wasted resources and degraded model performance. Unfortunately, today's collective communication libraries operate as black boxes, hiding critical information needed for effective root cause analysis.

We propose Mycroft, a lightweight distributed tracing and root cause analysis system designed to address previously hidden reliability issues in collective communication. Mycroft's key idea is to trace collective communication states and leverage internal control and data dependencies to resolve reliability problems in LLM training. Mycroft has been deployed at ByteDance for over six months to debug collective communication–related issues at runtime. It detected anomalies within 15 seconds in 90% of cases and identified the root cause within 20 seconds in 60% of cases. We also conducted extensive fault injection experiments to demonstrate Mycroft's capability and efficiency.

## 1   Introduction

Deep learning has achieved significant breakthroughs, particularly in large language models (LLMs). State-of-the-art LLMs are efficiently trained with various parallelism strategies, including data parallelism (DP) [36, 52, 53], pipeline parallelism (PP) [16, 35, 36, 51], and tensor parallelism (TP) [53, 70]. Models are typically trained on tens of thousands of GPUs for days or weeks [14, 20, 37, 55], interconnected through collective communications (Coll) over Remote Direct Memory Access (RDMA) networks. These setups use NICs with bandwidths up to 400 Gbps to achieve high throughput, low overhead network transmissions.

The high demand for training efficiency is accompanied by reliability challenges. For example, Meta reported over 105 restarts across more than 100 VMs, averaging 1.25 incidents per day and impacting 61,000 GPU hours [31]. Various hardware or software failures or performance problems may occur throughout the training process, causing crashes or slowdowns that hurt training efficiency and model progress.

Although tracing, monitoring, and anomaly detection methods have been proposed to address reliability issues [3, 4, 6, 15, 61, 63], real-time diagnosis remains challenging. Especially when problematic layers lack observability, training jobs may suffer from silent timeouts, fail-slows, or performance degradations [3, 6, 11, 15].

We argue that the performance and reliability of LLM training rely heavily on Collective Communication Libraries (CCLs), which manage distributed data transmission over networking layers. These communications follow complex control and data flows defined by the training frameworks, introducing critical dependencies across multiple levels. However, today's CCLs function as black boxes, lacking observability into critical internal states. As a result, when reliability issues occur, it is difficult for system developers and operators to identify root causes, resolve failures or performance issues, and resume training - leading to wasted hardware resources and human labor.

In this paper, we demonstrate the importance of collective communication level (Coll-level) observability, which exposes fine-grained dependencies from both flows and data chunks, providing meaningful insights for analyzing the root causes of failures or performance anomalies. However, enhancing observability at the Coll level is challenging because it requires real-time analysis and meaningful trace data collection with minimal overhead.

We build Mycroft, a distributed tracing and root cause analysis system that addresses previously hidden reliability issues at the Coll level in LLM training. Mycroft performs fine-grained tracing through lightweight instrumentation, capturing the necessary dependencies to expose communication states. To our knowledge, it is the first system to leverage Coll-level dependencies to debug LLM training reliability issues. Mycroft achieves efficient runtime diagnosis through its trigger mechanism and dependency-driven root cause analysis.

We evaluate Mycroft on a testbed with 32 NVIDIA A100 GPUs. The results show that Mycroft incurs lower overhead than state-of-the-art reliability systems, even those lacking Coll-level observability. To assess Mycroft's anomaly detection and root cause localization capabilities, we conduct extensive fault injection experiments covering common hardware and software issues. Mycroft successfully detects anomalies and locates the corresponding root causes.

---

[#]This work was completed while Yangtao Deng was an intern at ByteDance.
[*]Yangtao Deng and Lei Zhang contributed equally to this paper.
[†]Lei Zhang and Zhuo Jiang are corresponding authors.

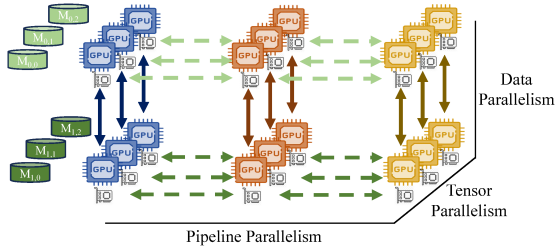

**Figure 1.** Hybrid parallelism in LLM training.

Mycroft has been deployed at ByteDance since October 2024. It achieves a 100% success rate in detecting Coll-level problems and detects anomalies within 15 seconds in 90% of cases. In 60% of cases, it identifies the root-cause GPU(s) within 20 seconds. We also provide real-world case studies demonstrating how Mycroft debugs complex Coll-level problems, and share our experience integrating Mycroft with other debugging systems in ByteDance's production environments.

## 2 Motivation

### 2.1 Reliability in LLM Training

Reliability is critical in LLM training for two reasons. First, large GPU clusters experience frequent anomalies - Microsoft reports a failure every 45 minutes [19], Meta observed 419 interruptions in 54 days during LLaMa 3 pre-training [7], and ByteDance sees at least one failure per day in its clusters [4]. Second, such anomalies cause significant delays and resource waste. Training relies on the complex hybrid parallelism, including DP, PP, and TP through collective communication across a large number of GPUs, as shown in Figure 1 [7, 20, 36, 69, 70]. Collective communication operations (CollOps) are desired to efficiently transmit Gigabytes of data within hundreds of milliseconds across GPUs. Rapid detection and mitigation can significantly improve the Effective Training Time Ratio (ETTR) [21], enhancing both training performance and resource efficiency [10, 35, 65].

In this paper, we focus on two typical types of anomalies.

- *Gray failures*, where training halts without immediate or obvious alerts [4, 20, 63], such as a `ReduceScatter` Op becoming stuck. In such cases, GPUs or NICs may remain idle until the problem is discovered and the job is restarted [58].
- *Performance issues*, also known as fail-slows [11, 15], stragglers[61], or performance degradations [3, 6], which slow down training without completely stopping it. These may appear as short-term fluctuations or long-term bottlenecks.

### 2.2 Problem: Lack of Observability

In LLM training, problems can manifest unpredictably at any point. When observability is limited in certain layers, developers often observe only high-level symptoms - such as software timeouts, exceptions, or degraded network throughput - which tend to appear uniformly across ranks or machines, making root cause diagnosis difficult.

For example, in a real-world failure where a GPU became defective during training, all ranks got stuck on network communications almost simultaneously, without raising an exception or output on any machine. The job eventually failed after 10 minutes when NCCL timeout was triggered. Unfortunately, there was no further evidence to investigate the faulty machine. Thus, various offline testing benchmarks were deployed for preliminary checks. The problem was not reproduced until GPT-2 training was run on multiple machines for more than six hours.

The core challenge in this case is the lack of proper observability. Since the anomaly occurs in a CollOp, observing fine-grained system states within the operation is necessary to reveal which software or hardware component causes the issue. Such traces in the CCLs also reveal how the anomaly propagates. This enables us to locate the culprit rank more accurately and determine the root cause. Thus, more targeted actions can be taken for mitigation, instead of replacing a group of machines and restarting until the anomaly occurs again.

### 2.3 Current Reliability Solutions

As shown in Table 1, several tracing and diagnosis systems exist for LLM training observability and anomaly detection. These tools are widely used for benchmarking, such as `perftest` [47], NCCL tests [41], cross-verification, and others [26, 63]. Together, they provide real-time or offline logs, which are primarily used in manual inspection. Here, we focus on PyTorch, RDMA networking, and GPU-related observability, which is at a relatively lower level in LLM training stacks.

**Op-level** refers to tracing at the granularity of each operation by instrumenting PyTorch's Coll API calls, *e.g.,* capturing the start time, end time, and overall count of processed operations. Kineto [30], part of the PyTorch Profiler [50], provides Op-level performance observability by capturing call stacks and integrating with tools such as Holistic Trace Analysis [29]. Chakra [56] formalizes workloads as execution graphs, capturing operations and timelines. Its execution traces are adopted by Mystique [24] for benchmarking and simulation. GREYHOUND [61] records CollOp timestamps for performance issues detection and segments them into iterations heuristically.

**Kernel-level** refers to the observability of CUDA libraries, typically represented by GPU kernel execution traces. Nsight

Table 1. Existing collective communication-related tracing tools.

| Tool | Granularity | RDMA Observability | GPU Observability | Gray Failure | Performance Issues | Distributed Analysis | Real time |
|---|---|---|---|---|---|---|---|
| Kineto [30] | Op-level | ✗ | ✗ | ✗ | ✗ | ✗ | ✗ |
| Chakra [56] | Op-level | ✗ | ✗ | ✗ | ✗ | ✗ | ✗ |
| Mystique [24] | Op-level | ✗ | ✗ | ✗ | ✗ | ✗ | ✗ |
| GREYHOUND [61] | Op-level | ✗ | ✗ | ✗ | ✓ | ✓ | ✗ |
| Nsight [42] | Kernel-level | ✗ | ✓ | ✓ | ✗ | ✗ | ✗ |
| NPKit [33] | Kernel-level | ✗ | ✓ | ✓ | ✓ | ✗ | ✗ |
| NVRx [45] | Kernel-level | ✗ | ✓ | ✓ | ✓ | ✓ | Minutes |
| XPUTIMER [3] | Kernel-level | ✗ | ✓ | ✓ | ✓ | ✓ | Minutes |
| Aegis [6] | RDMA-level | ✓ | ✗ | ✓ | ✓ | ✓ | ✗ |
| **Mycroft** | Coll-level | ✓ | ✓ | ✓ | ✓ | ✓ | Seconds |

Systems [42] provides comprehensive system-wide visualization and analysis across CPUs, GPUs, and other components. NPKit [33] traces events from NCCL [44], RCCL [2], and MSCCL [32], including delays in CollOps and kernel executions. Nvidia-resiliency-ext (NVRx) [45] gathers traces from CUDA Profiling Tools Interface (CUPTI) [39] and torch.distributed [48]. XPUTIMER [3] is a real-time diagnostic framework that instruments key Python API calls and critical kernels at the C++ level.

**RDMA-level** refers to recording RDMA-related statistics or metrics for a CollOp. Aegis [6] records the index, duration, and throughput of work requests in an individual CollOp to characterize RDMA behaviors. However, these metrics only indicate whether the RNIC works normally. As a result, they fail to precisely track anomaly propagation across GPUs.

**Limitations.** The observability tools described above are limited in their effectiveness and coverage at runtime. For Op-level and RDMA-level methods, a fundamental drawback is the lack of fine-grained observability. If a CollOp is stuck, these systems neither generate event traces nor expose the internal states of a CollOp for further investigation. Kernel-level methods typically rely on CUDA-level tracing that can capture detailed execution information for every GPU event. This level of granularity comes at a high performance cost. In most cases, we only need selective tracing near the problem area. There is no need to capture the entire CUDA execution timeline. For example, Nsight Systems can only be used on a single server; NPKit caused a two-thirds drop in bus bandwidth in our experiments; and NVRx relies on a low sampling rate and provides results every 60 seconds.

Moreover, many of the above methods generate event traces that fail to reveal the internal states of problematic CollOps. Leveraging only CUDA-level or Op-level tracing is insufficient to reveal executions on other hardware and system layers. These tools also commonly lack a distributed view; they collect traces only on the local machine and often provide limited analysis capabilities. The diagnostic process remains labor-intensive, rendering debugging inefficient.

### 2.4 Challenges

Achieving meaningful observability at Coll level introduces several challenges. We highlight the key ones that are critical for building an efficient system to debug LLM training problems at runtime.

**Fine-grained observability.** The diagnostic solution must trace precisely at the layer where the problem manifests; without this, debugging becomes slow and inefficient.

**Low overhead and scalability.** Tracing must provide sufficient time granularity and scale while minimizing overhead from instrumentation, storage, CPU usage, and network load, otherwise negatively impact the training performance (e.g. in [3], traces generated from Python or C++ APIs can exceed 100MB per iteration per GPU).

**Distributed analysis.** A typical LLM training task may involve up to 1000 machines, each generating massive volumes of data. Correlating information across machines to determine the root cause is usually time-consuming, and even pinpointing faulty GPUs across just two machines has taken several minutes in recent work [3]. Automatic and efficient collection and analysis from a distributed perspective are thus essential.

## 3 Understanding Dependencies in CCL

### 3.1 Dependencies

We dive deeper into the intricacies of dependencies in collective communications. Specifically, we outline the types of dependencies necessary to identify the root causes of failures or performance issues in LLM training.

**Intra-node dependencies.** From a single node's perspective, executing a CollOp in CCLs involves tight hardware-software co-design among the CPU, GPU, NIC, PCIe, and other components. This tight coupling makes the control loop complex and potentially asynchronous. Data chunks are typically split across multiple network paths. Each path transmits its part independently, and the completion is determined only after all paths have finished transmission. Each

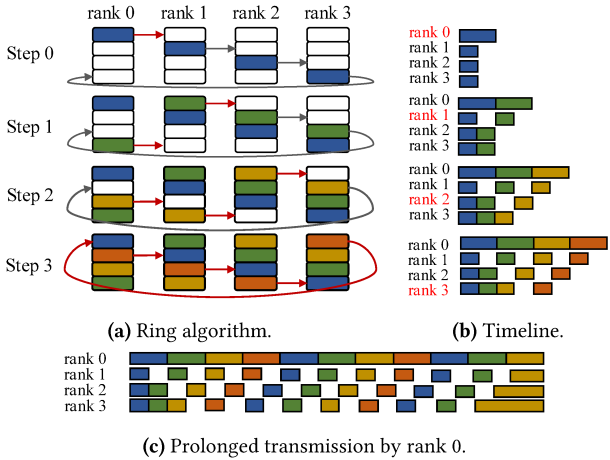

**(a)** Ring algorithm. **(b)** Timeline.

**(c)** Prolonged transmission by rank 0.

**Figure 2.** Dependencies in an `AllGather` slowdown.

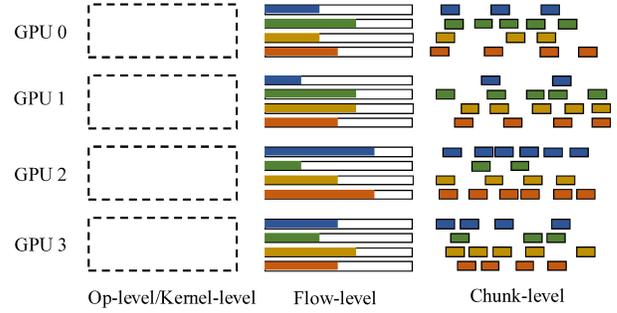

**Figure 3.** Trace granularity comparison. Chunk-level traces also contain system states as arguments.

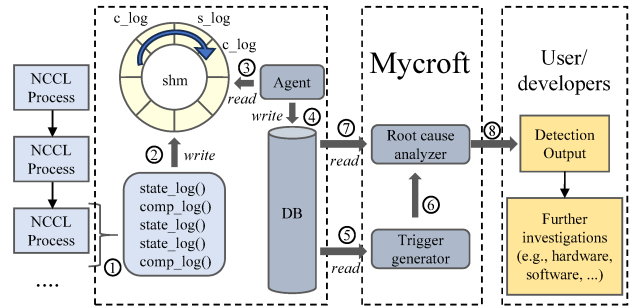

**Figure 4.** Mycroft Architecture

path may involve copying data between different memory or storage media. If multiple components from heterogeneous hardware are involved (*e.g.,* CUDA kernels and CPU processes), additional coordination mechanisms are needed. For example, a kernel may update its progress as it prepares data for transmission, while a CPU proxy process continuously polls for ready data to send via RDMA. From a debugging perspective, certain components are often the source of anomalies. In these cases, observing the system states during data transmission is crucial. It means tracing the progress of each component in the control flow, as the performance of downstream components depends largely on their upstream counterparts.

**Inter-node dependencies.** From a global perspective, where multiple nodes transmit data using different parallelism mechanisms, numerous data dependencies exist among nodes in collective communication.

In many Coll primitives, data is synchronized within the same communication group, where downstream components wait for upstream data to arrive before continuing. This synchronization often occurs at a fine granularity, involving chunks split into smaller slices. In this process, a single straggler node can degrade the performance of the entire CollOps in the group. Figure 2 shows an example of `AllGather` Op across four ranks, where rank 0 consistently transmits data chunks more slowly than the others. In step 0, all ranks except rank 0 complete as expected. In step 1, rank 1 is delayed, waiting for data from rank 0, which in turn delays rank 2 in step 2, cascading the slowdown through all ranks. Ultimately, all ranks complete together at the synchronization barrier, with the slowdown cascading across all ranks in the group.

Beyond synchronization within a single communication group, ML training often involves multiple forms of parallelism with interdependencies. For example, at the start of an iteration, ranks in a DP group collect updated weights. Afterward, they exchange activations or gradients with their PP neighbors. During the linear algebra phase, TP operations generate intra-host traffic. Once each mini-batch completes its matrix computations, ranks aggregate gradients within each DP group. In such nested control paths, any failure or slowdown in one Coll group can impact subsequent CollOps across the entire system.

Although the above slowdown process is illustrated with a ring topology, the same principle applies to tree-based CollOps where relatively complex data synchronization mechanisms are defined. In a tree topology, data progress propagates hierarchically from parent nodes to child nodes (or vice versa). Each node must wait for its upstream parent to finish sending before forwarding to its downstream children. A slowdown at any node in the upper levels of the tree therefore delays all dependent subtrees, often amplifying the impact compared to a ring topology. This constitutes a hierarchical dependency that must be considered for further diagnosis.

**Computation-communication dependencies.** CollOps can be slowed by factors such as resource contention. CollOps such as `ReduceScatter` and `AllReduce` involve additional kernel executions to perform `Reduce`. Under parallelism strategies [59], these communication processes can be delayed when kernel execution overlaps occur on the same node. In such cases, `Reduce` kernels are launched but remain unscheduled for unexpectedly long periods.

Table 2. Categories and trace metrics.

| Categories | Metrics |
|---|---|
| Metadata | IP, comm_id, Gid, GPU_id, Channel_id, QP_id |
| Operation | timestamps, op_name, op_seq, msg_size |
| Chunk | stuck_time, total_chunks, GPU_ready, RDMA_transmitted, RDMA_done |

## 3.2 Tracing Coll-level Dependencies

Existing Op-level and Kernel-level methods lack the fine-grained observability described above, motivating the need for more detailed tracing to support runtime debugging. Thus, we introduce the idea of Coll-level tracing, consisting of the following two aspects.

**Flow-level.** Since CCLs involve multiple network flows that construct the overall communication topology, we record traces at the granularity of each network flow (*e.g.,* from QPs and other network paths). This is critical because many existing monitoring tools aggregate the behaviors of all network flows as one operation. However, flows within the same CollOp are not bound to the same NICs and paths. In LLM training, network topologies are in fact defined at the Flow level instead of the Op level, which is not observable in previous coarse-grained methods. As a result, occasional traffic spikes, jitter, or congestion that affect only one flow cannot be detected or localized by Op-level tracing.

**Chunk-level** tracing monitors the transmission of each chunk — the smallest data unit per network path in CollOps, typically a few megabytes in size. Instead of logging the start and end timestamps of each chunk, which incurs huge data overhead similar to Kernel-level tracing, we track system states that reveal training health. We implement Chunk-level tracing by aggregating execution data from hardware (CPU, GPU) and software (*e.g.,* RDMA protocol) components. It allows us to quantify their operational states by the amount of data operated during a CollOp in each short time window. Our approach enables precise identification of underperforming components without incurring the overhead of tracking every chunk event.

Figure 3 illustrates our selected fine-grained tracing level. Compared to Op-level and Kernel-level tracing, Flow-level and Chunk-level traces reveal internal system states when a CollOp gets stuck, and provide more granular performance measurements than recording only the execution time of a single CollOp.

## 4 Mycroft Design

We built Mycroft, a system that enables real-time gray failure and performance issue detection and root cause analysis. Figure 4 represents the overall system design. We introduce key system components to illustrate how Mycroft improves reliability for LLM training at scale.

### 4.1 Opportunities

We first summarize our observations on real-world reliability issues, highlight critical opportunities, and explain the rationale for our design choices.

**Cluster-wide impact via inter-node dependencies.** From dependency analysis and real-world observations, we find that reliability and performance issues originating from a single node can propagate through inter-node dependencies, ultimately affecting the entire cluster.

**Rapid anomaly spread.** CollOps over RDMA networks are optimized for high-throughput, low-latency transmission. For example, transmitting a few gigabytes of data typically take tens of milliseconds. When such a CollOp stalls, it quickly blocks the entire group of ranks, which in turn blocks subsequent nested communication groups. Previous studies [3, 4, 61] and our experiments in Section 7.2 show that these anomalies can propagate cluster-wide within a few hundred milliseconds.

**Revealing the global state machine with minimal data.** As discussed in Section 3, while capturing critical path information across hardware and system components is essential, only a small amount of data is needed to reveal problematic system states. In practice, this observability can be achieved primarily by recording run-time states in CCLs rather than collecting extensive traces such as CUDA kernel executions.

### 4.2 Instrumentation

We instrument NCCL[1] to record critical runtime states from the NCCL proxy threads, capturing how data transmission progresses on each device.

**NCCL hardware – software coordination.** During initialization, NCCL creates multiple network flows (called "channels") between each rank pair, each containing several QPs. Data is split into small chunks and evenly distributed across these flows. When transmission begins, a CUDA kernel and a CPU-side NCCL proxy thread cooperate in sending data. The kernel copies a chunk from GPU memory to preallocated GPU buffer, after which the NCCL proxy transmits the chunk via RDMA to the receiver and polls for a Completion Queue Entry (CQE) to mark completion. This coordination is asynchronous as discussed in Section 3.

**Tracepoints.** Mycroft provides two types of tracepoints to instrument NCCL's critical path - `completion log` and `real-time state log`- which collect key runtime metrics that reveal data transmission progress. Table 2 summarizes the per-log fields collected on each rank.

- A `completion log` records when a CollOp finishes and includes metadata such as start and end timestamps, bytes transmitted, local and remote NIC information, and other CollOp metadata.

---
[1]Although our implementation is based on NCCL [44], Mycroft is a general solution to all CCL variants. We further discuss this in Section 9.

**Algorithm 1:** Trigger Mechanism

**Input:** Sampled IP list $ip\_list$, time $t$
**Output:** Active abnormal trigger
1 **Function** Trigger($ip\_list, t$):
2     $log \leftarrow$ Acquire($ip\_list, t - \Delta, t$);
3     **if** *no Coll Ops completed in log* **then**
4        **return** *failure trigger, abnormal ip*;
5     **else**
6        **if** *throughput drops by half or Coll Op interval doubles in log* **then**
7           **return** *straggler trigger, abnormal ip*;
8        update normal throughput and Coll Op interval;

- A `real-time state log` is generated periodically (e.g. every 100ms) while a CollOp is in progress, capturing the accumulated progress during that interval. It reports data transmission progress across all devices, covering Stream Multiprocessor (SM) copies and RDMA writes. `real-time state log` is generated in each window until the CollOp completes or the NCCL proxy thread exits or crashes.

**Data Collection.** Mycroft preallocates a fixed-size circular buffer in the host machine's shared memory space during NCCL's initialization stage. At each tracepoint, it acquires a pointer to the next available buffer slot and writes trace data directly into the memory space, minimizing the control loop overhead. This runs on NCCL's internal streams to support asynchronous measurement, adding virtually no overhead to the critical path. A separate read-only agent consumes the data and uploads it to a cloud database via Kafka, avoiding locks and preventing back pressure. This design enables continuous runtime tracing with negligible overhead.

### 4.3 Real-time Trigger Mechanism

Mycroft 's always-on analysis backend processes the collected trace data. The main challenge is handling the massive volume of traces generated across up to tens of thousands of GPUs in a single training job, while still enabling fast detection and analysis at scale.

Mycroft implements a rapid active trigger mechanism by **sampling a subset of ranks across the cluster and monitoring all CollOps**, reducing both data volume and analysis time. This benefits from our observation that a failure or performance issue quickly cascades to the whole cluster, thus abnormal executions can be quickly detected from any sampled rank.

In practice, Mycroft samples at least one rank per DP group to ensure coverage in common LLM parallelism architectures. Even without topology information, other sampling schemes remain effective since anomalies propagate quickly. Additionally, to ensure scalability, Mycroft limits sampling to at most 10 ranks. This mechanism allows Mycroft to remain effective at the scale of tens of thousands of GPUs, as always-on detection is required only on a small set of sampled nodes.

**Algorithm 2:** Dependency-driven Analysis

**Input:** Abnormal IP $ip$, Time $t$
**Output:** Suspected Root Cause
1 **Function** AnalyzeFailureRootCause($ip, t$):
2     $last\_log \leftarrow$ AcquireGroupLastLog($ip, t$);
3     $(MinOp\_rank, rank\_log) \leftarrow$ CheckMinOp($last\_log$);
4     **if** *MinOp_rank exists* **then**
5        $failure\_category \leftarrow$ CheckRCTable($rank\_log$);
6        **return** ($MinOp\_rank, failure\_category$);
7     **else**
8        $(MinData\_rank, rank\_log) \leftarrow$ CheckMinData($last\_log$);
9        $failure\_category \leftarrow$ CheckRCTable($rank\_log$);
10        **return** ($MinData\_rank, failure\_category$);
11 **Function** AnalyzeStragglerRootCause($ip, t$):
12     $GL \leftarrow$ AcquireGroupLogs($ip, t$);
13     **for** $(rank\_log, rank\_id) \in GL$ **do**
14        update iteration start and end times;;
15        document $rank\_ids$;
16     **for** $rank\_id \in rank\_ids$ **do**
17        **if** *constant late starts or ends* **then**
18           **return** ($rank\_id, straggler\_category$);
19     **return** None;

Algorithm 1 describes Mycroft's trigger rules. To detect failures or stuck issues, Mycroft checks if the sampled rank stalls mid-operation with `real-time state log` but without producing `completion log`. Mycroft detects performance issues if *its throughput drops by half or the operation interval doubles*. These heuristics are based on practical runtime experience, and the rules are configurable across different model performance patterns (we further discuss heuristics tuning in Section 9). A successful trigger output identifies only the suspicious time point, without necessarily pinpointing the specific range of affected machines.

## 5 Debugging

Once a potential failure or performance issue is triggered, Mycroft begins identifying its likely root cause of the problem at the specific time point $t$.

### 5.1 Distributed State Machine

Mycroft determines the root cause using a global state machine built from instrumentation data capturing the dependencies. The state machine is constructed from logs within a short time window $(t - \Delta, t)$, which contains either the last active system state for a failure or stall, or the execution details of a performance issue. Because problems cascade properly, $\Delta$ is small by design, limiting the trace data volume and ensuring analysis of the overall job at scale.

Table 3. Root cause analysis principles.

| Level | Problem | Rule |
|---|---|---|
| Chunk-level | Failure | Each rank should transmit the same amount of data. |
| Chunk-level | Performance | Each component should finish within expected execution time. |
| Chunk-level | Performance | Each component should not block the downstream ones. |
| Flow-level | Failure | Each flow should complete. |
| Flow-level | Performance | Each flow should take similar execution time. |
| Flow-level | Performance | Each flow should start and end at similar time. |

## 5.2 Root Cause Analysis Rules

Algorithm 2 presents Mycroft's root cause analysis process. Mycroft narrows down the root cause of a potential problem step by step, starting with the identification of the affected Coll group. For failures or stalls, there is typically a single responsible group whose stuck rank propagate problems to other Coll groups through inter-node dependencies.

After locating the affected Coll group, Mycroft further determines the specific responsible components. Table 3 summarizes the analysis principles derived from fine-grained traces at different levels, based on observed dependency violations in collective communication.

For failures or stalls, Mycroft pinpoints the network flow that either fails to complete transmission or makes the least progress, as indicated by the latest system states. These network flows are then identified as the root cause and aggregated accordingly. For example, when all flows of a rank are stuck, the rank is likely the root cause.

For performance issues, Mycroft evaluates both chunk-level and flow-level execution against expected behavior. At the chunk level, it checks whether each component finishes within the expected execution time and whether it blocks any downstream component. At the flow level, it verifies that all flows take similar execution times and start and finish at roughly the same time. Any component or flow that violates these rules is marked as a potential performance bottleneck. This process also involves heuristics such as configurable thresholds to detect violations. For example, to define stragglers that start or end later than other normal nodes, we choose one second as the threshold from experience. We discuss heuristic accuracy in Section 9.

## 5.3 Locate Intra-node Root Causes

Mycroft assists in narrowing down potential directions for investigation by examining system states along the critical path to determine root causes. Table 4 summarizes representative root causes that can be revealed by analyzing hardware–software coordination in data transmission. For instance, if transmission stalls because some data chunks are prepared by the GPU but never transmitted, this suggests

Table 4. Potential local and remote root causes.

| State | Condition | Local cause | Remote cause |
|---|---|---|---|
| Not started | ①=②=③=0 | Uninitialized | Blocked |
| Not transmitted | ①>② | RDMA issue | Receiver not ready |
| Not delivered | ②>③ | RDMA issue | Receiver failed |
| GPU not ready | ①=②=③>0 | GPU issue | - |

Note: ①, ②, ③ represents number of chunks for GPU_ready, RDMA_transmitted, and RDMA_done stage. Multiple conditions can be met simultaneously.

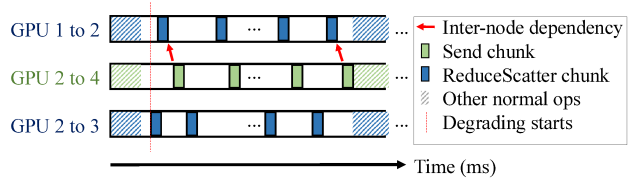

Figure 5. Fine-grained Coll-level trace for a straggler case: prolonged CollOp progress visibility. GPUs 1, 2, and 3 are in a DP group, while GPUs 2 and 4 are in a PP group.

intra-host network issues. Conversely, if all GPU-prepared data is successfully transmitted, the RDMA network is likely functioning correctly and the issue may lie instead in the GPU or software layer.

In some cases, observing system states from a single rank at a specific time point can be insufficient for debugging. Temporally, analyzing a rank's states over time can reveal recurring patterns. For example, if we frequently observe ①=②=③, this indicates a likely GPU-related issue. Spatially, comparing states between ranks helps diagnose failed or slow communications. For example, if a transmission is stalled with the sender reporting ①=②>③ while the receiver reports ①=②=③, then the failure is likely attributable to the receiver.

## 5.4 Runtime Example

Figure 5 illustrates a runtime example of how Mycroft detects a GPU problem as the root cause. GPUs 1, 2, and 3 form a DP group, and GPU 1 is significantly slower than the others. Due to inter-rank dependencies, this slowdown cascades to GPUs 2 and 3 within the DP group. Meanwhile, GPUs 2 and 4 belong to a PP group, so the delay on GPU 2 also propagates to GPU 4. When Mycroft processes the trace data, it first identifies the affected communication groups from abnormal execution times — detecting both the DP group (1–2–3) and the PP group (2–4) as impacted. It then analyzes execution progress within each group, pinpointing GPU 1 as the rank that consistently executes kernel copies slowly and blocks overall communication progress. Since all other delays stem from this single GPU through dependency chains, Mycroft attributes GPU 1 as the root cause. Mycroft locates the faulty GPU with no human involvement, and the GPU is rapidly replaced to resume the training job.

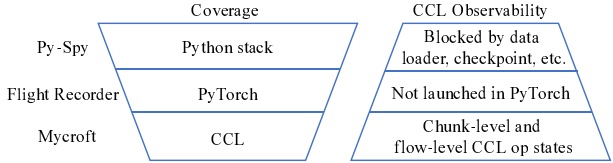

**Figure 6.** Mycroft integrates with other reliability systems.

## 6 Implementation

### 6.1 Implementation and Deployment

We instrument NCCL 2.21.5 with 1100 lines of C++ code, including tracing APIs for managing efficient in-memory trace data as a typical tracing system, along with additional tracepoints that capture metrics from the NCCL proxy's data transmission critical path as discussed in Section 4. In practice, we only instrumented less than 10 tracepoints which requires minimal engineering effort.

Mycroft's tracepoints generate trace data in local shared memory within each worker's container, occupying a fixed 512MB on each host. An agent on each host asynchronously reads and uploads data to a cloud database. This pipeline introduces no significant lag in our production environment.

We build Mycroft's backend with 4000 lines of Python code. It efficiently queries trace data from the cloud database and analyzes it by applying triggering and root cause analysis rules discussed earlier. Mycroft is deployed on a single server, using fewer than 64 CPU cores and under 100GB of memory, as an always-on system to monitor ByteDance's LLM training jobs with up to tens of thousands of GPUs.

The cloud database serves as a caching layer for two main purposes. First, it allows Mycroft to access data from the recent past on demand when a proactive trigger occurs, ensuring that essential information is not lost or irretrievable. Second, Mycroft temporarily retains this data to handle complex scenarios where abnormalities arise but do not match predefined detection rules. In such cases, the cached data is used either as an observability tool to construct and present execution traces, or to compute ad-hoc statistics for specialized analysis beyond Mycroft 's standard capabilities. Based on our production experience, a training job utilizing 10,000 GPUs generates approximately 3 TB of data per day. This data is retained for one day before being discarded.

### 6.2 Integration with Other Debugging Systems

In addition to the above failures and performance issues, the distributed training can get stuck due to other reasons like data starvation or synchronization. To reduce false positives, Mycroft provides interfaces for external passive trigger mechanisms on top of its active trigger mechanism for real-time detection. As shown in Figure 6, Mycroft will dump Python stacks of each rank using py-spy to identify dataloader or checkpoint stuck and dump PyTorch CollOps collected by Flight Recorder to identify the synchronization issues.

**py-spy** [1] is a sampling profiler for Python programs. It gets call stacks of the program without modifying the code. We use it extensively to dump out Python stacks when training tasks become idle, in which we check whether the threads stuck in a particular function. Firstly, stack traces of all relevant Python programs are dumped automatically when Mycroft detects a trigger. Stacks are then grouped by process commands across all GPUs and mapped into a topology grid, where each block represents a GPU rank with its current Python call stack. In this grid, identical call stacks will be marked in the same color. This is particularly useful as stuck threads typically have different call stacks from the rest, making them stand out on the grid for ease of troubleshooting.

**Flight Recorder** [49], as provided in PyTorch's recent versions, stores the traces of the latest $N$ CollOps in a ring buffer. Each CollOp trace includes the CollOp ID, the sizes of input and output tensors, execution state, and communication process group ID. CollOp traces in the ring buffer can be automatically dumped, similar to the process of automatically dumping Python stack traces using py-spy. We extract CollOp stacks to analyze the rank synchronization issue by figuring out the device with last operation on each CUDA stream in a process group, and aggregating all the trace stacks to visualize and identify abnormal devices. For example, we can find the stuck GPU that does not launch any CollOp, deadlock of communication groups, or the different input and output size of some ranks from others.

**Integration.** In practice, we integrate these three reliability systems to provide a complete picture of networking reliability for LLM training. These systems complement each other to bound the problematic layer of reliability issues. For example, if py-spy indicates some dataloader synchronization problems, even though Flight Recorder and Mycroft may also detect problems, it is easier to locate the real root cause. If py-spy and Flight Recorder both cannot detect the gray failures or performance issues, the problem lies inside the CCL layer, and Mycroft works on exposing the root cause. We provide more real-world case studies in Section 8.

## 7 Evaluation

We evaluate Mycroft and answer the following questions:

- Can Mycroft detect and localize diverse faults in collective communication?
- How does Mycroft compare to existing systems, regarding capability and performance?
- How does Mycroft perform in real-world LLM training production environments?

### 7.1 Mycroft capability

To thoroughly evaluate Mycroft 's ability to detect root causes, we conduct fault injection experiments using seven representative fault types drawn from production experience, including resource contention, congestion, and network hardware issues. These types cover most known anomalies in our experience. As shown in Figure 7, we inject anomalies into a single machine.

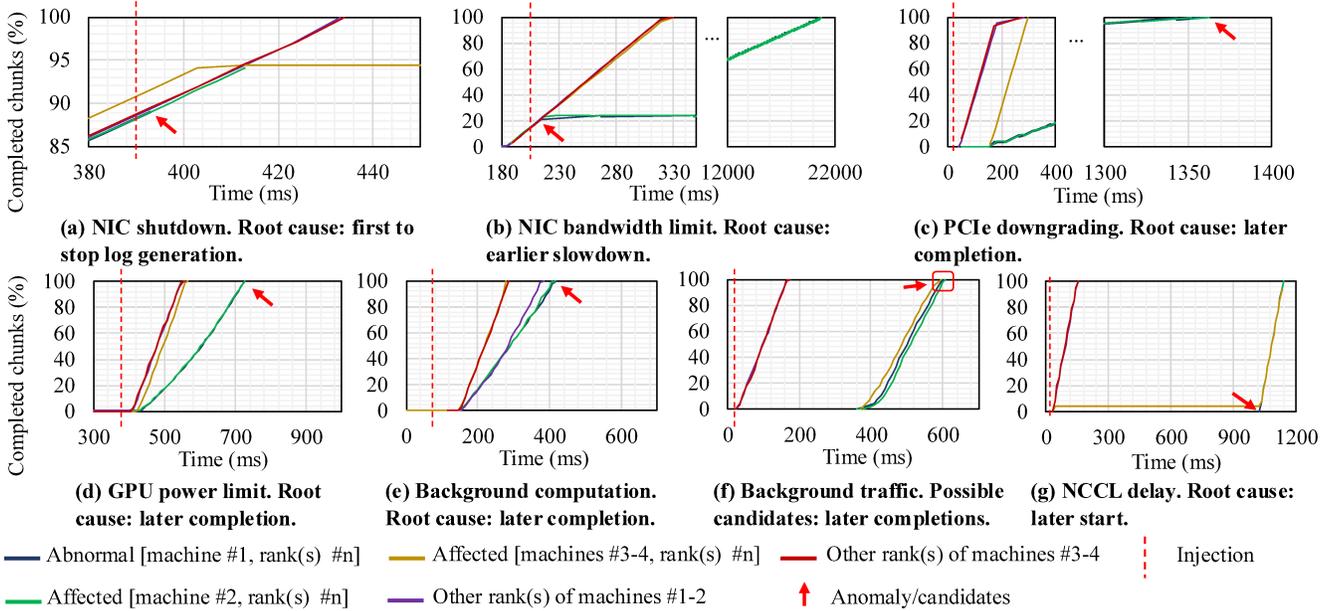

**Figure 7.** Operation progress from Coll-level traces after an injection in machine #1 during `ReduceScatter`. Placing machines #1-2 in PP stage 1 and machines #3-4 in PP stage 2. Affected peer ranks stop transmission or log generation in (a), slow down their transmission in (b)-(f), and wait for transmission in (g). Other ranks finish `ReduceScatter` normally.

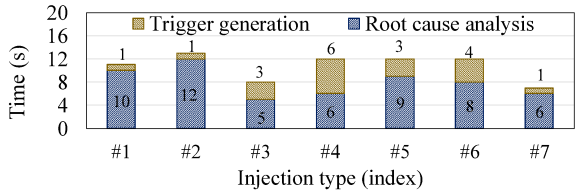

**Figure 8.** Time to detect root causes for the injections.

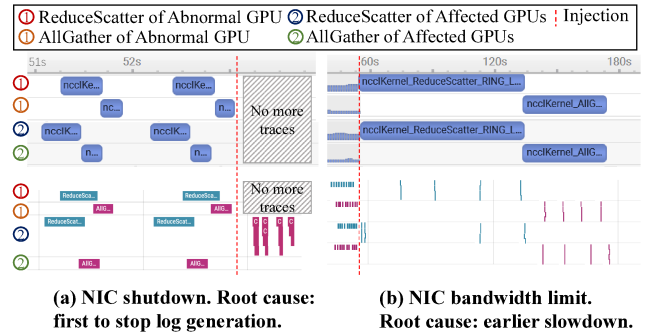

**Figure 9.** Nsight traces (above) VS Mycroft traces (below) after NIC down or NIC bandwidth limit injection: no difference for all GPUs in Nsight.

**Testbed and workload.** We run experiments on 32 NVIDIA A100-80GB-SXM4 GPUs [43] across four machines. The intra-host connections use NVLink and PCIe. The inter-host connections rely on four NVIDIA ConnectX-6 RNICs [38] of each machine. We choose GPT with Megatron-LM [53] as the workload, where we configure hybrid parallelism (TP=8, PP=2, DP=2). The tests run with CUDA 12.2 and Torch 2.1.0.

**NIC shutdown (#1).** The abnormal rank is detected as it stops earlier than the others in the same DP group, although both stop generating traces shortly. Ranks on machines #3 and #4 stop communication due to missing data from the first PP stage. The remaining ranks complete `ReduceScatter` normally and stall at the next `AllGather`, waiting for the synchronization. This experiment shows how communication and computation dependencies drive anomaly propagation.

**NIC bandwidth limit (#2).** Similar patterns appear after the injection. The slowed DP group continues at a reduced speed. By checking the temporary progress when they start to slow, and ensuring that `GPU_ready > RDMA_transmitted`, we are able to locate the root cause in the NIC.

**PCIe downgrading (#3) and GPU power limit (#4).** Both injections affect the intra-host computation. CollOps in the abnormal DP groups are delayed. Although peer ranks in the second PP stage communicate faster, they start later due to dependencies of the unfinished data transmission from the last operation. Since computation is slower, the abnormal rank is found by comparing completion timestamps. The abnormal rank always finishes last in the DP group.

**Background computation (#5).** Extra GPU load is added to machine #1 in this injection. Similar to the above two cases,

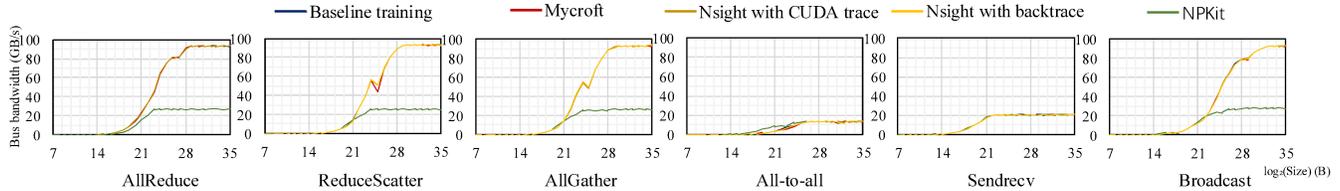

**Figure 10.** Impact on NCCL Tests on 32 A100 GPUs. Nearly overlapping lines for baseline, Mycroft, and Nsight indicate minimal overhead. NPKit's low bandwidth shows significant overhead.

computation becomes the bottleneck. Since the injection affects all eight ranks on machine #1 [40], eight DP groups are impacted. All ranks in the first PP stage slow down, while `ReduceScatter` in the second PP stage completes faster. Unlike earlier cases, all ranks begin data transmission almost simultaneously, as the second PP stage ranks are peers of the slowed ranks. Still, we can identify the culprits on machine #1 by comparing their completion times.

**Background traffic (#6).** We inject extra NIC traffic on machines #1 and #3 using perftest. Each NIC handles the transmission of one or two GPUs exclusively, based on the NCCL topology. As shown in Figure 7, unaffected ranks complete normally. However, ranks on machines #1 and #3, and those in their DP groups slow down. A small completion timing gap helps identify the culprit. Mycroft also finds that `GPU_ready` always equals to `RDMA_transmitted`, but `RDMA_done` is usually smaller. It indicates delayed acknowledgments and confirms the network as the root cause.

**NCCL delay (#7).** We inject delays by introducing a one-second sleep in NCCL's proxy (`sendProxyProgress`) with a fixed probability (p=0.000001). The proxy, as a CPU-side thread handling GPU communication, can impact the progress of CollOps when delayed. As shown in Figure 7, other ranks complete the CollOps normally, while ranks within the affected DP groups in both PP stages experience significant delays. In this case, the abnormal rank initiates communication at the very end, while its peers have already transmitted portions of data to neighboring ranks (step 0 in Figure 2). The late start helps us detect the culprit.

**Efficiency.** We break down the latency from the start of an anomaly to the alert made by Mycroft for each injection experiment. In Figure 8, the latency is no more than 13 seconds overall. Most of the latency is contributed by the centralized root cause analysis on the controller. Nevertheless, it is still prompt compared to the minute-level alerts of other baselines [3, 45]. The responsiveness of Mycroft comes from its continuous monitoring of the sampled machine and the effective log selection in the algorithm design.

### 7.2 Comparison of Capability

We compare Mycroft's diagnostic capability with Nsight [42], chosen for its wide adoption and kernel-level tracing support. Other tools either perform similarly to Nsight while only reporting final diagnosis results (*e.g.,* NVRx [45], XPUTIMER [3]), or are restricted to NCCL Tests (*e.g.,* NPKit [33]). We run the NIC shutdown and bandwidth limit injections experiments as above and the same workload.

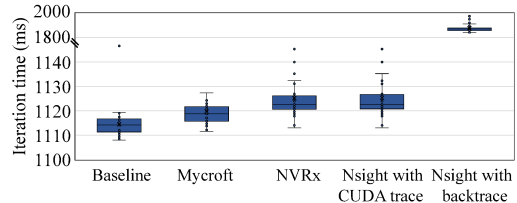

**Figure 11.** Impact on the iteration time for GPT training with Megatron. Nsight with CUDA trace only tracks kernel timestamps.

Figure 9 shows the results, using traces aligned offline from four machines. Nsight records the start and end timestamps of CUDA kernels, treating each kernel as a single unit per GPU. In the NIC shutdown case, Nsight fails to trace the affected `ReduceScatter` kernels on all GPUs, making it impossible to identify the failed machine. In fact, ranks from unrelated groups finish normally, while peers on two neighboring machines stall during `ReduceScatter` from Mycroft's traces. By analyzing which rank stops log generation first and comparing the GPU and NIC buffer states, Mycroft detects the root cause. In the bandwidth limit scenario, Nsight reports all the ranks from the task delay their `ReduceScatter` almost simultaneously, including those unaffected. In Mycroft's detection, the abnormal rank will slow down earlier than other affected ranks. In other words, the rank that always starts transmission later is detected as abnormal by Mycroft here. These results indicate that Kernel-level tools lack not only the support for real-time analysis but also the granularity and visibility to reveal fine-grained dependencies.

### 7.3 Comparison of Overhead

We compare Mycroft's overhead with NPKit [33], Nsight Systems [42], and NVRx [45]. In this experiment, we use the same testing environments as above, and use both GPT with Megatron and NCCL Tests as workload.

We first compare NCCL-test bus bandwidth and GPT iteration time before and after deploying Mycroft or other tools separately on four machines. For Nsight Systems, NVRx, and NPKit, we leverage their tracing capabilities in CUDA libraries and kernels. Nsight supports an extra CUDA API backtrace capability. As shown in Figure 10, under lightweight NCCL Tests workload, Mycroft and Nsight incur overheads similar to the baseline, while NPKit adds much higher overhead, especially in CollOps, reducing bandwidth to about one-third.

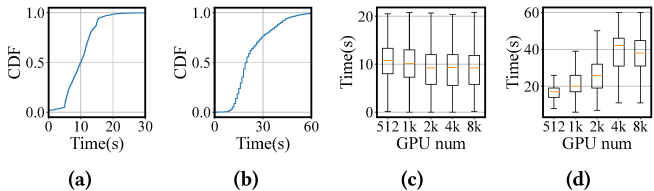

**Figure 12.** Trigger time and root cause analysis time shown as CDFs (a-b) and comparisons across different GPU counts (c-d), measured in ByteDance's production environment.

**Table 5.** CPU & memory usage in Megatron's GPT training.

| Metric | Idle | Baseline | Mycroft | NVRx | Nsight with CUDA trace | Nsight with backtrace |
| --- | --- | --- | --- | --- | --- | --- |
| CPU | 1.5% | 10.7% | 11.0% | 11.5% | 11.2% | 16.3% |
| Memory | 1.48% | 1.82% | 1.85% | 1.85% | 2.41% | 2.43% |

In Figure 11, we choose Nsight and NVRx for comparison since only these two commercial tools support GPT training with Megatron. The average iteration time increases marginally from 1116 ms to 1119 ms after the deployment of Mycroft, showing a degradation of less than 1%. NVRx extends the time to 1124 ms. 1125 ms is needed for Nsight with CUDA trace under its default settings. All of them degrade the performance by less than 1%, except Nsight in its backtrace. The time surges to 1900 ms due to excessive CUDA-lib tracepoints and the call stack. These results confirm that Mycroft introduces minimal interference to the original training tasks.

Table 5 shows the CPU and memory overhead during training. Mycroft does not increase CPU utilization. NVRx consumes more CPU cycles, but its result analysis takes minutes in our experiment and only reports at a predestined interval. Nsight backtrace introduces approximately 50% more overhead. Memory usage especially varies a lot due to log generation and code instrumentation. By checking the storage, Mycroft generates only approximately 46.8 KB of trace data per iteration per machine. Even at the scale of 1,200 machines, the size remains around 186 KB per machine per iteration (about 3 seconds). In contrast, Nsight generates 15 MB per iteration in our experiment even only with CUDA tracing. XPUTIMER produces up to 0.78MB of logs per GPU when training a Llama-70B model on 16 A100 GPUs [3]. This is mainly due to their excessive tracing granularity, resulting in large volumes of logs or unfiltered tracepoints.

### 7.4 Mycroft in Production

Mycroft has been deployed in ByteDance's production systems since October 2024, and keeps monitoring large-scale LLM training jobs with more than 128 GPUs. Simultaneously, there are up to 20 concurrent training jobs.

**Detection.** We summarize the online diagnosis performance during November and December 2024. Mycroft detects 13221 interruptions from all training jobs. Note that similar to other works [3, 5, 6], it is difficult to label the root cause of all anomalies in production environment, which makes it hard to determine Mycroft's accuracy like false positive rate. However, Mycroft maintains a 100% training job interruption detection rate by cross-checking all known abnormal jobs. Although not all of these interruptions are collective communication related, Mycroft processes 1253 times root cause analysis and provides a list of suspicious GPUs. Among that, Mycroft locates only one problematic network flow in 705 cases.

**Scalability and efficiency.** In ByteDance's daily monitoring tasks, Mycroft also provides problem detection and root cause analysis with high efficiency and scalability. Figure 12 represents the detection and root cause analysis time distributions. We define trigger time as the elapsed time between the true onset of an anomaly and the moment our system issues a trigger. Mycroft runs detection every 10 seconds on each training job, so trigger time consists of the detection interval delay and processing latency. We define root cause analysis time as the time between a trigger is issued and an anomaly rank is identified. As shown in Figure 12a, Mycroft achieves 90% stuck detection in 15 seconds, and as in Figure 12b, Mycroft achieves all root cause analysis within a minute, where about 60% tasks are done within 20 seconds. We further evaluate scalability across different training scales. As shown in Figure 12c, Mycroft maintains stable trigger time regardless of scale. The processing time for root-cause analysis increases with scale (Figure 12d), with the majority of the overhead attributable to retrieving data from the database over the past window (10 seconds by default). Scaling beyond 10,000 GPUs may necessitate additional optimizations, such as decentralization or other advanced engineering techniques, which we leave for future work. Nevertheless, Mycroft has already demonstrated the ability to complete analysis within a minute across all current scenarios.

## 8 Case Studies

Beyond the example provided in Section 5.4, we demonstrate other cases from the production environment to show Mycroft's ability for dependency revealing and its flexibility to integrate with other reliability tools.

**Case One - Straggler propagation due to inter-node dependency.** In this case, Mycroft detected a performance drop in an FSDP [69] training job, where `AllGather` occasionally ran slower than expected. Runtime traces revealed that one of the four NCCL channels was slow, taking 75 ms to finish compared to 30 ms on the others. Chunk-level system states further pinpointed the faulty GPU, which delayed the data preparation in the middle of the CollOp and caused slowdown on all downstream GPUs in the same DP group. Mycroft not only detected the faulty GPU as the root cause at runtime, but also provided flow-level insights that ruled out networking hardware (since the other three network paths were normal) and instead pointed to abnormal SM behavior. This gave system developers clearer guidance for deeper debugging.

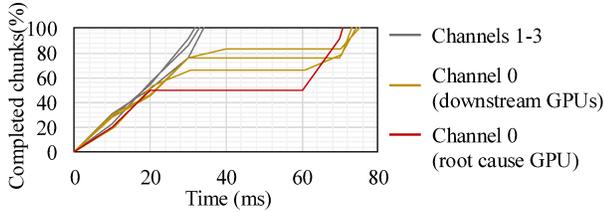

**Figure 13.** Chunk-level and flow-level dependency exposing root cause GPU.

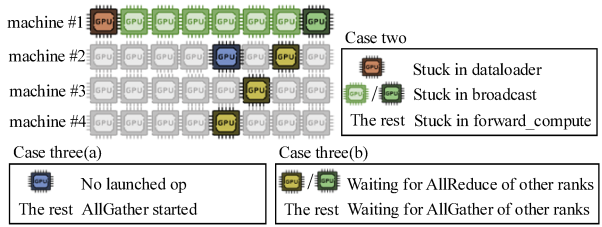

**Figure 14.** Stack traces from py-spy and synchronization issues from Flight Recorder.

**Case two - Guidelines for py-spy to diagnose stuck threads.** In this case, a large-scale training task became idle during the running stage. With the output dependency information as guidelines, we used py-spy to further identify which nodes might have stuck threads. py-spy is instrumented with a sampling rate for all training tasks. We manually collected stack traces of all train_megatron processes from machines of potential anomalies and grouped those that had identical call stacks. The traces and the parameters were scrutinized offline. In Figure 14, stacks are divided into three groups. The call stacks for the deep green and light green ranks were blocked in `broadcast`, the red one was stuck in `dataloader`, while other ranks were stuck at `forward_step` (megatron/schedules.py). This helps us to pinpoint the faulty red rank quickly as it failed to load data for TP while its TP peers were waiting for the data in `broadcast`.

**Case three - Guidelines for Flight Recorder to diagnose stuck ranks.** Alerts from Mycroft were also used with Flight Recorder for further analysis. Case three(a) in Figure 14 shows that a blue rank has not launched any CollOp, but other ranks launched `AllGather` and waited for the blue rank. We could identify that the GPU of the blue rank had a breakdown with Flight Recorder. Case three(b) in Figure 14 shows a deadlock case due to framework misconfigurations of users, where each rank had two process groups and the CollOps were deadlocked. The yellow and dark green ranks started the `AllReduce` op and waited for other ranks whose `AllReduce` op states were not scheduled, because they were waiting for the `AllGather`.

## 9 Discussion

**Generality.** Mycroft's generality lies not only in the idea of tracing dependencies in CCLs but also in the system itself. As discussed in Section 6, Mycroft's tracing library and analyzing system are designed to efficiently collect, query, and process CollOp metrics. To adapt Mycroft to other CCL varieties or hardware stacks, system developers can redefine the specific metrics along the data transmission's critical path, adjust the tracepoints, and design detection and analysis rules accordingly. To clarify, NCCL is among the most complex CCLs and involves more heterogeneous hardware stacks than many other scenarios in today's LLM training, and metrics listed in Table 2 are general for tracking system states in collective communication. For example, if data transmission in a CCL does not involve GPU SM copy, then Mycroft doesn't collect such metrics and the root cause analysis is just simplified.

**Real-time capability.** Despite generating massive amounts of trace data, Mycroft maintains real-time monitoring by distributing trigger generation across sampled machines and selectively processing logs in root cause analysis. Other tools either rely on coarse-grained data that hampers responsiveness (*e.g.*, 30-second counters [4] and minute-level timeouts in PyTorch or RDMA alerts), or cannot process fine-grained data in time. Thus, they can only serve as auxiliary tools for Mycroft.

**Handling asymmetric collectives.** Although Mycroft primarily targets Ring-based Coll issues, it naturally extends to imbalanced communication patterns such as MoE workloads, All-to-All, and M-to-N collectives. Mycroft tracks critical system states along with each CollOp with no difference, and the same for detecting whether a transmission is intrinsically slow. A greater challenge occurs when a GPU is stalled because other nodes have not yet reached synchronization or barrier points in earlier steps, introducing substantial noise into detection. To mitigate false positives, it is essential to disambiguate such synchronization-induced delays by incorporating the model's ground truth execution semantics.

**Tuning heuristics and parameters in analysis.** Defining anomalies in LLM training is challenging, especially when issues cause subtle performance drops rather than failures. For performance issues, Mycroft takes operation latency and network throughput as the major metrics in collective communication, while allowing new heuristics and parameters. For example, bandwidth drops to 80% may appear anomalous but can result from overlapping communication, leading to false positives; we therefore use a 50% bandwidth reduction as the trigger. Another example is GPU-level straggler detection, where threshold selection directly affects accuracy: overly strict thresholds increase false positives (e.g., detecting master rank which is expected to handle heavier workloads), whereas overly relaxed thresholds reduce precision. In practice, we set the threshold at 1 second to determine if a rank is a straggler—that is, if it starts or finishes more than 1 second later than the others. With these heuristics, Mycroft successfully identified performance issues among many different training jobs that vary in model, scale, and hardware. We leave automatic threshold adjustments as future work.

**Lessons.** To implement instrumentation in NCCL, we chose to add tracepoints directly to NCCL's software stack rather than relying solely on NCCL plugins. This approach was necessary because we aimed to capture control path information; NCCL plugins can only provide networking-related details and lack access to workload-specific data. This may not apply to newer NCCL versions or to other collective communication libraries that allow passing control information through plugins.

**Limitations.** While Mycroft can automatically detect faulty nodes without human intervention, its observability is confined to collective communication, limiting full end-to-end visibility in training jobs. Operating solely at the communication layer, it cannot capture application-level intent, such as intentional overlap between communication and computation or designed parallelism. As a result, distinguishing true anomalies from benign variations remains challenging in complex topologies. Deeper root causes, such as underlying hardware faults or software bugs, often still require complementary tools or manual diagnostics. For example, in the debugging case of Section 5, once Mycroft identified the problematic GPU, offline hardware checks confirmed that memory was not the source of failure. We believe tighter integration with other diagnostic tools could enable more automated identification of such deep-root causes, and we consider this as future work.

## 10 Related Work

**Real-time monitoring.** Real-time host-side tools monitor machine states using existing counters during the task lifecycle. Minder [4], Autopilot [17], and other approaches [67] operate based on out-of-band metrics, such as GPU utility and memory usage. These methods rely heavily on sampling rates and cannot capture communication patterns from the tasks themselves. KungFu [28] monitors gradient metrics using user-defined policies to enable adaptive training. However, it requires a custom library and cannot track communication patterns. Real-time network-side tools usually include ping-like measurements. Pingmesh [12] and P-pingmesh [25] monitor end-to-end latencies between arbitrary pairs of servers periodically. AdapCC [68] profiles traffic latency and throughput in NVLinks [46] and inter-host networks (RDMA and TCP), but its out-of-band traces do not explicitly reveal the communication process with a low sampling rate.

**Fault-tolerance.** Redundancy-based approaches, such as Malleus [23], ReCycle [9], Bamboo [57], and Oobleck [18], focus on fast recovery rather than detecting performance or fault degradation. They do not provide fine-grained tracing for communication analysis. CheckFreq [34] dynamically adjusts the checkpointing frequency to minimize fault recovery overhead. Gemini [60] introduces checkpointing traffic scheduling across different storage tiers. MegaScale [20] mitigates storage bottlenecks through data sharing among the corresponding servers. FlexRR [13] combines a flexible synchronization with dynamic peer-to-peer re-assignment to address performance issues. These works focus on optimizing other infrastructures rather than communication.

**Dependency and causal analysis.** Dependency analysis has been widely explored beyond LLM training. ExChain [22] targets programming languages by monitoring exceptions and tracking causal links. In distributed systems, tools like Dapper [54], X-Trace [8], Pivot Tracing [27], and Hindsight [66] analyze event dependencies to monitor request execution, while Chain-of-Event [64] and Cloud Atlas [62] build causal graphs from system telemetry and documentation. Compared with those systems, Mycroft is specifically designed for LLM scenarios.

## 11 Conclusion

In this paper, we highlight the importance of tracing collective communication states and leveraging internal control and data dependencies to address reliability problems in LLM training. We present Mycroft, a lightweight system for diagnosing distributed training anomalies. Through targeted instrumentation in CCLs, sampled trigger generation, and root cause analysis, Mycroft reveals fine-grained flow-level dependencies while minimizing real-time overhead for efficient tracing and analysis. Mycroft reduces end-to-end detection time to the second level, providing actionable dependency-driven insights for developers.